\documentclass[preprint,showpacs,preprintnumbers,amsmath,amssymb]{revtex4}

\usepackage{graphicx}% Include figure files
\usepackage{dcolumn}% Align table columns on decimal point
\usepackage{bm}% bold math

\usepackage{graphicx,color,psfrag}

\newcommand{\e}{\varepsilon}

\newcommand{\ds}{\displaystyle}
\newcommand{\be}{\begin{equation}}
\newcommand{\ee}{\end{equation}}
\newcommand{\bea}{\begin{eqnarray}}
\newcommand{\eea}{\end{eqnarray}}
\newcommand{\bi}{\begin{itemize}}
\newcommand{\ei}{\end{itemize}}

\begin{document}
%\doublespacing

\title{\textbf{SURFACE-MEDIATED NON-LINEAR OPTICAL EFFECTS \\IN LIQUID CRYSTALS}}

\author{V.O. Kubytskyi}
\email{kubba@univ.kiev.ua}
\author{V.Y. Reshetnyak}%
\affiliation{%
Physics Faculty,  Kyiv National Taras Shevchenko University,\\
Prospekt Glushkova 2, Kyiv 03022, Ukraine\\}%

\author{T.J. Sluckin}
\affiliation{School of Mathematics, University of Southampton,\\ Southampton SO17 1BJ, UK }%

\author{S.J. Cox}
\affiliation{School of
Engineering Science, University of Southampton,\\ Southampton SO17
1BJ, UK}%

\date{\today}

\begin{abstract}
We make a phenomenological model of optical two-beam interaction
in a model planar liquid crystal cell. The liquid crystal is subject
to homeotropic anchoring at the cell walls,  is surrounded by thin
photosensitive layers, and is subject to a variable potential across
the cell. These systems are often known as liquid crystal
photorefractive systems. The interference
between the two obliquely incident beams causes a time-independent
periodic modulation in electric field intensity in the direction
transverse to the cell normal. Our model includes this field
phenomenologically by affecting the potential at the walls of the
cell. The transverse periodic surface potential causes spatially
periodic departures from a pure homeotropic texture. The texture
modulation acts as a grating for the incident light. The incident
light is both directly transmitted and also subject to
diffraction. The first diffracted beams result in energy exchange
between the beams. We find that the degree of energy exchange can
be strongly sensitive to the mean angle of incidence, the angle
between the beams, and the imposed potential across the cell. We
use the model to speculate about what factors optimize non-linear
optical interaction in liquid crystalline photorefractive systems.
\end{abstract}

\maketitle

%\newpage

\section{Introduction}

It has long been known that nematic liquid crystals act as non-linear
optical media \cite{N1}. The electric fields in a strong light beam reorient the
liquid crystal director. In so doing they affect the dielectric properties
of the medium, and hence its light transmission and reflection. Thus the
liquid crystal reacts differently to a high intensity beam than when the
intensity is low. A slab of liquid crystal exhibits analogous properties
when irradiated by two beams rather than by a single beam. The liquid
crystal responds to the interference pattern between the beams. The result
is a grating in the liquid crystal cell, which diffracts the incoming beams.
The lowest order diffracted beams from each incident beam act to reinforce
the other, leading to the phenomenon of beam amplification. This latter
phenomenon is the subject of this paper.

In general beam-coupling is a highly non-linear phenomenon and as such
requires intense beams to manifest itself. However, it has been found
experimentally that there are circumstances when beam-coupling appears to
occur at much lower light intensities. The device possibilities of these
high beam-coupling conditions have meant that these systems have attracted
much interest. Two particular interesting systems obtain when either the
liquid crystal is doped by dye molecules, or when the liquid crystal cell is
sandwiched between walls consisting of photoconducting material. Now free 
charges can play an important role in the non-linear optics -- in the
former case the ions move inside the liquid crystal itself, whereas in
the latter case they affect the boundary conditions to which the liquid
crystal is subject. In this paper we consider the second of these cases --
that in which the liquid crystal is surrounded by photosensitive layers. An
associated feature of such systems is that the degree of beam-coupling is
strongly dependent on, and amplified by, a low-frequency voltage across the
liquid crystal cell.

The existence of ions in these systems has led to this phenomenon being
linked with photorefraction \cite{N2}. In photorefractive systems moving charges
lead to non-linear optical effects. These so-called photorefractive liquid
crystal systems exhibit large optical non-linearities, at least partly
because two non-linear optical processes seem to be manifesting themselves
simultaneously. This statement, however, while true, is insufficient even to
give the most basic description of the physics of beam-coupling in these
systems. In this paper we present a phenomenological model, which although
by no means a complete description of the system, provides a basic
understanding of some of the most striking features of the experiments. The
most important of these features is the observation that the largest
beam-coupling effects occur when the grating period is comparable to the
cell thickness.

Experimental work in photorefraction in liquid crystals dates back about
decade. In the cases of interest in this paper, the effect results because a
spatially modulated light field causes a modulation of the electric field
either in the aligning layer itself \cite{N3,N4}, \cite{N5,N6,N7} or in the interface
between the LC and the aligning layer \cite{N8,N9,N10}. In the first case the liquid
crystal cell is lined by photoconductive aligning layers, whose electrical
resistance is decreased by light irradiation. This increases the electric
field in the liquid crystal bulk, which in turn causes a spatially modulated
reorientation of the director in the cell. The effect is reversible; the
induced gratings disappear when the incident light is switched off. By
contrast, in the second case the photorefraction is controlled by the
processes in the interface between LC and aligning surfaces. Both of these
layers may be nominally insensitive to light. The resultant spatially
modulated electric field induces a reorientation of the director in the bulk
and a permanent grating.

Theoretical work on these systems has concentrated on extending existing
photorefractive concepts, which have been developed for optically isotropic
systems. In this case, the key inputs into an experiment are: the cell
thickness $L$, the light wave-length $\lambda $, the grating period $\Lambda $,
and the dielectric constant $\varepsilon $ of the isotropic medium. Kogelnik
\cite{N11} developed a coupled-wave theory which can predict the response of
volume holograms (i.e. thick gratings). Klein \cite{N12} gave the criterion for a
grating to be thick in terms of the parameter $Q=2\pi L\lambda /\Lambda
^2\sqrt \varepsilon $. The coupled wave theory begins to give good results
when $Q\ge 10$. Montemezzani and Zgonik \cite{N13} have extended the Kogelnik
coupled-wave theory to the case of moderately absorbing thick anisotropic
materials with grating vector and medium boundaries arbitrary oriented with
respect to the main axes of the optical indicatrix. The dielectric tensor
modulation takes the form
\be
\hat {\varepsilon }=\left[ {\varepsilon _r^0 +\varepsilon _r^1
\cos \left( {K\cdot r} \right)} \right]+i\left[ {\varepsilon _i^0
+\varepsilon _i^1 \cos \left( {K\cdot r+\phi } \right)} \right].
\ee  Galstyan  \textit{et al} \cite{galstyan} have also presented a variant of this idea  
applied to anisotropic thin holographic media.

The key extra piece of physics in liquid crystal cells is that
the director is anchored by the cell walls. As a result the
spatial modulation of the dielectric function is considerably more
complicated than the Montemezzani-Zgonik form. In addition, the
liquid crystal cell parameters are often in the so-called
Raman-Nath regime \cite{N14,raman_nath}, which corresponds to thin gratings.
For thin isotropic gratings with a one-dimensional
refractive index modulation, the theory is well developed (see for
example \cite{N14}). For such a system, for example, Kojima
\cite{N16} used a phase function method to understand the
diffraction problem for weakly inhomogeneous anisotropic materials
in the Raman-Nath regime, assuming a dielectric function spatial
modulation $\varepsilon \sim \cos \left( {\Omega t-Kx} \right)$.

In this paper we shall study the diffraction and energy transfer of two
light beams intersecting in a nematic liquid crystal cell with strong
homeotropic anchoring at the cell walls. The beam-coupling can be amplified
by a DC-electric field, which is applied to the cell perpendicular to the
cell walls (\textit{Oz}-direction). This problem corresponds to that addressed 
experimentally by Korneichuk \textit{et al} \cite{korneichuk}.

The presence of the two beams causes a periodic
lattice in the light intensity field in the cell bulk and its boundaries. In
addition, the laterally periodic light intensity also causes a modulation in
the \textit{dc-}electric field potential at the cell boundaries. This paper
addresses the photorefraction problem phenomenologically. 

Specifically, we consider here a restricted problem with two major \textit{caveats. } Firstly
we shall not examine too closely the origin and mechanism of this
modulation. We simply remark that it can and does result from different
physico-chemical phenomena taking place at the cell walls. Likewise in this
simple approach, we shall suppose that the physics of the electric field in
the liquid crystal is driven by dielectric processes, and that charge
transport does not play a major role in determining director orientation or
light scattering. Elsewhere we shall relax both of these constraints.

The key to understanding beam-coupling in these systems lies in the
following observation. The surface potential modulation produces a spatially
modulated electric field. The resulting torque on the liquid crystal
director distorts the initial homogeneous homeotropic alignment. The
consequence is an anisotropic medium with a spatially modulated director and
hence optical axis. The test beam -- or the beams that write the grating --
diffract from the liquid crystal cell, which now possesses a spatially
modulated refractive index. One may then calculate beam diffraction and
inter-beam energy transfer.

The paper is organized as follows. In Section II we determine the electric
field profile in a cell subject to a light-induced periodic modulation of
the surface potential. In Section III we calculate the director distribution
inside the liquid crystal cell subject to this spatially modulated electric
field. Then in Section IV we present results of calculations of beam
diffraction and energy transfer. Finally in Section V we present some brief
conclusions, and focus on possible extensions of the model.

\begin{figure}[ht]
    \centering
        \includegraphics[width=0.45\textwidth]{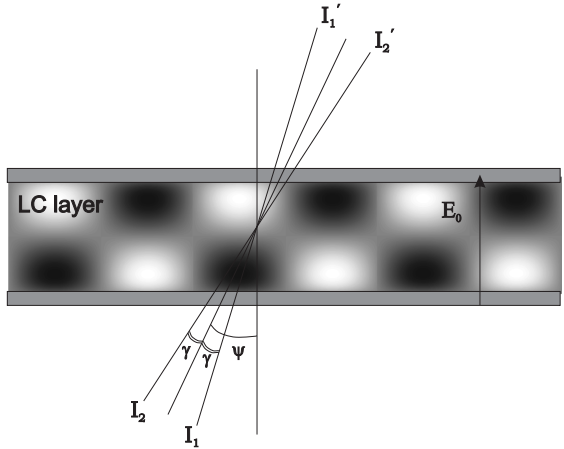}
        \caption{Schematic picture of a two-beam coupling experiment, showing
meaning of quantities used in the paper.}
    \label{fig:process}
\end{figure}

\section{Electric field within the \protect\\liquid crystal slab}

We consider two equal frequency light beams with
wave numbers $\mathbf{k}_1 
\approx \mathbf{k}_2$ inside the medium. These 
beams give rise to electric fields $E_1 ,\;E_2
$, and intensities $I_1 ,\;I_2 $ proportional to the squares of
the respective electric fields:

\bea
\mathbf{E}_1 &= & E_{10}\exp \left( {i \mathbf{k_1}\cdot \mathbf{r}} \right),\\
\mathbf{E_2} &=& E_{20} \exp \left( {i \mathbf{k_2} \cdot \mathbf{r}} \right),\\
\mathbf{k_1} &=& \left( {k\sin \left(
{\psi +\gamma } \right),0,k\cos \left( {\psi +\gamma } \right)}
\right), \\
\mathbf{k_2} &=& \left( {k\sin \left(
{\psi -\gamma } \right),0,k\cos \left( {\psi -\gamma } \right)}
\right).
\label{B1} 
\eea

The bisector of the beams makes an angle $\psi $ with the cell normal and
$2\gamma <<\psi $ is the angle between the beams. Initially we suppose that
the scattering by the cell is weak, and thus the light transmission through
the cell is close to unity.

The beams interfere in the liquid crystal slab, forming a complex intensity pattern with wave number $q=k_{1x} -k_{2x} $. In principle the pattern must be calculates self-consistently. In practice it may be possible to treat surface-induced and bulk-induced effects separately. In this section we concentrate on the particular effect at the surfaces. At the bottom (i.e. incident) interface, the interference pattern of light takes the form:

\bea
&I( {x,z=0})=I_1 +I_2 +2\sum (I_1 I_2)^{1/2} \cos qx, \nonumber \\
&q=k_{1x} -k_{2x}
\label{B2}
\eea

Likewise, at the top substrate we have an analogous pattern, but shifted in
phase with respect to the lower substrate:

\bea
&I\left( {x,z=L} \right)=I_1 +I_2 +2\sum (I_1 I_2)^{1/2} \cos (qx+\delta ), \nonumber \\
&q=k_{1x} -k_{2x}
\label{B3}
\eea

\be
\delta =\left( {k_{1z} -k_{2z} } \right)L
\label{eq:delta}
\ee
In the absence of the light beams, we suppose a voltage $\Phi _0 $ across
the liquid crystal cell. This is a key input to the theory. In
photorefractive systems the optical non-linear effects are large and
strongly amplified by a voltage across the cell. In the simple theory
presented here the non-linear effect in the absence of an external field is
strictly zero.

We now make the hypothesis that the spatial distribution of light intensity
induces a modulation in the surface potentials. The effect of the light
beams is to modify these potentials slightly. The boundary conditions on the
electric potential at the top and bottom substrates can now be written:

\bea
 \varphi \left( {z=0,x} \right)&=&-\Phi _0 /2+\Phi _1 \cos qx,\; \\
 \varphi \left( {z=L,x} \right)&=&\Phi _0 /2+\Phi _1 \cos \left( {qx+\delta }
\right)
 \label{B4}
\eea where $\Phi _1 =\alpha (I_1 I_2)^{1/2}$. The parameter
$\alpha$ is a phenomenological quantity, and in principle is
different for each surface configuration. Here we are supposing
that the surface preparation of the upper and lower surfaces is
identical. The parameter $\alpha $ can in principle be determined
independently by a Frederiks experiment \footnote{We might imagine a system in which the boundary conditions were
homogeneous, for example, and only one surface was attached to a
photosensitive layer. Only one of the surfaces would then be affected by an
incident light beam. A single light beam would then affect the effective
voltage across the cell. The effect would be proportional to the light
intensity, and the constant of proportionality would be the parameter
$\sigma $. The magnitude of the effect could be measured by monitoring the
critical Frederiks field.}.

We now proceed to determine the electric field potential within the liquid
crystal slab. The electric field obeys the equation

\be
\nabla \cdot {\rm {\bf D}}=0
\label{B5}
\ee
with $D_i =\varepsilon _{ij} E_j =\left[ {\varepsilon _\bot \delta _{ij}
+\left( {\varepsilon _{\vert \vert } -\varepsilon _\bot } \right)n_i n_j }
\right]E_j $, where $\bf n$ is the nematic director.

 We can solve eq.(\ref{B5}) using relation ${\rm {\bf E}}=-\nabla
\varphi $. At this stage we note that the equations for $\varphi$ and $\mathbf{n}$ must be solved self-consistently. However the
 liquid crystal is subject to homeotropic boundary conditions, and hence
except at very strong light intensities the director is closely aligned to
the direction perpendicular to the slab: $\bf n\approx \bf e_{z}$, and then
\be \label{eq:Laplace}
 \e_{\bot} \frac{\partial ^2 \varphi}{\partial x^2} + \e_{\|} \frac{\partial ^2 \varphi}{\partial z^2} = 0.
\ee

The problem to be solved is thus eq.(\ref{eq:Laplace}), subject to
the boundary conditions eq.(\ref{B4}). This problem can be solved
analytically, yielding
\begin{widetext}
\begin{equation}
	\varphi \left( {x,z} \right)=\Phi _0 \frac{z-L/2}{L}+\Phi _1 \left[ {\left\{
	{\cosh \left( {\tilde{q}z} \right)+\frac{\cos \delta -\cosh \left( {\tilde{q}L}
	\right)}{\sinh \tilde{q}L}\sinh \left( {\tilde{q}z} \right)} \right\}\cos qx-\frac{\sin
	\delta }{\sinh \tilde{q}L}\sinh \left( {\tilde{q}z} \right)\sin qx} \right],
	\label{B7}
\end{equation}
\end{widetext}
where $\tilde{q} = \sqrt{\frac{\e_{\bot}}{\e_{\|}}} q$.
The potential in the liquid crystal slab consists of the externally imposed
voltage plus a contribution linear in the surface perturbation induced by
the light-beam interference. This perturbation has the same periodicity in
the direction in the cell plane as the initial perturbation. However, the
behavior is complex as a result of the competing effects of the out-of-phase
surface perturbations.

The resulting electric field, in addition to the imposed external field
$E_0$ normal to the cell, has components in both the $x$ and $z$
directions. The contributions in the $x$ direction are particularly
important, because they lead to director distortion, and thus to refractive
index modulation. It will be this refractive index modulation which induces
the beam-coupling which we seek to describe.

The electric field inside the cell bulk is given by taking the gradient of eq.(\ref{B7}). We find:

\bea
 E_x \left( x,z \right)&=&E_{1q} \left( z \right)\cos qx+E_{2q} \left( z
\right)\sin qx \nonumber \\
 E_z \left( {x,z} \right)&=&E_0 +{E}'_z \label{B8}  \\
 {E}'_z \left( {x,z} \right)&=&E_{3q} (z)\cos qx+E_{4q} \left( z \right)\sin
qx \nonumber
\eea

where
\be
E_0 =-\frac{\Phi _0 }{L},
\ee

\be
E_{1q} =q\Phi _1 \frac{\sin \delta }{\sinh \tilde{q}L}\sinh \tilde{q}z,
\ee
\be
E_{2q} =q\Phi _1 \left( {\cosh \tilde{q}z+\frac{\cos \delta -\cosh \tilde{q}L}{\sinh
\tilde{q}L}\sinh \tilde{q}z} \right)
\label{B9}
\ee
\be
E_{3q} =-\tilde{q}\Phi _1 \left( {\sinh \tilde{q}z+\frac{\cos \delta -\cosh \tilde{q}L}{\sinh
\tilde{q}L}\cosh \tilde{q}z} \right)
\ee

\be
E_{4q} =\tilde{q}\Phi _1 \frac{\sin \delta }{\sinh \tilde{q}L}\cosh \tilde{q}z
\ee
It will be useful later to normalize the field $E_{iq}$ with respect to the externally imposed field:
\be
e_{iq} \left( z \right)=\frac{E_{iq} }{E_0 }=a_i \cosh \tilde{q}z+b_i \sinh \tilde{q}z,
\label{B10}
\ee
where the quantities $a_i ,b_i $ are given by:
\be
 \arrayrulewidth = 0.05em
\begin{array}{|c|c|c|}
        \hline
        i&  a_i & b_i \\
        \hline
        &&\\
        1&  0 & {\ds -q L \frac{\Phi _1}{\Phi _0} \frac{\sin \delta }{\sinh \tilde{q}L}}  \nonumber \\
        &&\\
        2&  {\ds -q L \frac{\Phi _1}{\Phi _0}} & {\ds -q L \frac{\Phi _1}{\Phi _0} \frac{\cos \delta -\cosh \tilde{q}L}{\sinh
\tilde{q}L}}  \nonumber \\
        &&\\
        3&  {\ds \tilde{q} L \frac{\Phi _1}{\Phi _0} \frac{\cos \delta -\cosh \tilde{q}L}{\sinh
\tilde{q}L}}& {\ds \tilde{q} L \frac{\Phi _1}{\Phi _0}}  \nonumber
\\
        &&\\
        4&  {\ds \tilde{q} L \frac{\Phi _1}{\Phi _0}\frac{\sin \delta }{\sinh \tilde{q}L} }& 0 \nonumber \\
        \hline
\end{array}
\ee

\section{Director profile in the liquid crystal cell }

We now determine the director profile in the liquid crystal cell in the
presence of the electric fields given by eqs.(\ref{B8}). The bulk free energy $F_V
$ of a distorted nematic liquid crystal in an applied electric field takes the form:
\begin{widetext}
\begin{equation}
	F_V =\frac{1}{2}K_{11} \int {\left( {\nabla \cdot \bf n}
	\right)^2dV+} \frac{1}{2}K_{22} \int {\left( {\bf n\cdot \nabla
	\times  \bf n} \right)^2dV+} \frac{1}{2}K_{33} \int {\left( {\bf
	n\times \nabla \times \bf n} \right)^2dV} -\frac{1}{2}\int {\bf
	D\cdot \bf{E} dV} \label{B11}
\end{equation}
\end{widetext}
with the electric displacement
$\bf D=\hat {\varepsilon } \bf E$, $\varepsilon_{ij} =\varepsilon
_\bot \delta_{ij} +\varepsilon _a n_i n_j $, with the anisotropic
part of the static dielectric constant $\varepsilon _a=\varepsilon
_{\vert \vert } -\varepsilon _\bot $.

 The electric field felt by the liquid crystal molecules has a number of
contributions. The first is the externally-imposed voltage. The second is
the periodic modulation in the $x$ direction discussed in the last section.
This is an indirect effect of the light field acting on the surface layer,
transmitted into the bulk as a result of the effect of the Laplace equation.
A final contribution comes from the direct effect of the light field on the
liquid crystal. We are assuming here that this can be neglected. The
justification for this is empirical, and derives from the observation that
in the absence of the surface layers, the effect essentially disappears
\cite{N18,N7}. The director field is now given by $n=\left( {\sin \theta \left(
{x,z} \right),0,\cos \theta \left( {x,z} \right)} \right)$, with $\theta $
small.

The variational problem to be solved consists of minimizing eq.(\ref{B11}),
subject to strong anchoring homeotropic boundary conditions $\theta \left(
{x,z=0} \right)=\theta \left( {x,z=L} \right)=0$ at each wall, and subject
also to the electric fields given in eq.(\ref{B8}). We simplify further by
supposing the so-called one constant approximation, i.e. the splay and bend
Frank-Oseen elastic coefficients are equal: $K_{11} =K_{33} =K$.

The relevant part of the thermodynamic functional eq.(\ref{B11}) is now given by:

\bea
F&=&\frac{1}{2} \int\!\!\!\!\int [ K[ (\theta_x^{'})^2+( \theta_z^{'})^2 ]- \nonumber \\
&&\varepsilon_a(( {E_x^2 -E_z^2 })\sin ^2\theta + E_x E_z \sin 2\theta)]dx dz.
\label{B12}
\eea
The Euler-Lagrange equation for this functional is:
\bea
&&K\left( {\frac{\partial ^2\theta }{\partial x^2}+\frac{\partial ^2\theta}{\partial z^2}} \right)+\nonumber \\
&&\varepsilon_a\left( {\left( {E_x^2 -E_z^2 }
\right)\sin \theta \cos \theta +E_x E_z \cos 2\theta } \right)=0.
\label{B13}
\eea
We solve eq.(\ref{B13}) in the limit $\dfrac{E_x }{E_0
}\leq 1,\quad \dfrac{{E}'_z }{E_0 }\leq 1$. This corresponds, roughly speaking, to 
high voltage or low beam intensities. Expanding to linear order in $\theta $
and $\dfrac{E_x }{E_0 }$, we obtain:
\be
\xi ^2\left( {\frac{\partial ^2\theta }{\partial x^2}+\frac{\partial
^2\theta }{\partial z^2}} \right)=\theta -\frac{E_x }{E_0 }
\label{eq1}
\ee
where the length scale $\xi $ is the relaxation length set by the bulk
electric field, with $\ds \xi ^{-2}=\frac{\varepsilon _a E_0^2 }{K}$. In order to do this, 
we linearize  eq.(\ref{eq1}) for components $\theta _{1q} =\vartheta _1
\left( z \right)\cos qx$ and $\theta _{2q} =\vartheta _2 \left( z
\right)\sin qx $ of the director reorientation of wave number $q$ in the cell plane.

It is convenient at this stage to reformulate the problem in terms of non-dimensional variables.
We define a rescaled length along the cell $\ds \rho =\frac{z}{L}$, $\rho \in [0,1]$, a rescaled transverse
wave vector $\mu =\tilde{q}L$, and a rescaled voltage $\ds \nu =\frac{L}{\xi } = L E_0(\frac{\e_a}{K})^{1/2}$. For some purposes it is convenient to measure length not from the plane of incidence, but rather from the mid-plane of the cell. We can define a length variable $\sigma = \rho -1/2$, and then inside the cell $\sigma \in [-1/2,1/2]$.

Eq.(\ref{eq1}) now reduces to

\be
\frac{d^2\vartheta _i \left( \rho \right)}{d\rho^2}-\kappa ^2\vartheta _i
\left( \rho \right)=-\nu ^2e_{iq} \left( \rho \right)
\label{B15}
\ee
with $\kappa ^2=\mu ^2+\nu ^2$.
 
This equation can be solved using standard methods and has solution:
\begin{widetext}
\bea
\theta (x,z)=-q L \frac{\Phi _1}{\Phi_0} \Bigl(\cos qx\sin \delta \left[ {\frac{\sinh \mu \rho }{\sinh \mu }-\frac{\sinh
\kappa \rho }{\sinh \kappa }} \right]  +\sin qx \Bigl[\cosh \mu \rho -\cosh \kappa \rho +\\ \nonumber
\frac{\cos \delta -\cosh \mu }{\sinh \mu }\sinh \mu \rho +\frac{\cosh
\kappa -\cos \delta }{\sinh \kappa }\sinh \kappa \rho \Bigr] \Bigr).
 \label{theta1}
\eea
\end{widetext}
For some purposes it is more convenient to rewrite eq.(\ref{theta1}) as:
\be \label{theta2}
\theta \left( {x,z} \right)=-q L \frac{\Phi _1}{\Phi_0}  \left[ {\begin{array}{l}
 \cos \frac{\delta }{2}\sin \left( {qx+\frac{\delta }{2}} \right)\left\{
{\frac{\cosh \mu \sigma }{\cosh \mu /2}-\frac{\cosh \kappa \sigma }{\cosh
\kappa /2}} \right\} \\
 \quad \quad \quad \quad \quad \quad \quad + \\
 \sin \frac{\delta }{2}\cos \left( {qx+\frac{\delta }{2}} \right)\left\{
{\frac{\sinh \mu \sigma }{\sinh \mu /2}-\frac{\sinh \kappa \sigma }{\sinh
\kappa /2}} \right\} 
 \end{array}} \right]\ee
The advantage of the expression (\ref{theta2}) is that the variation of $\theta(x,z)$ 
is expressed in terms of components that are respectively  out-of-phase  and  in-phase  with  the total  optical field intensities on the mid-plane. This expression explicitly exhibits the 
symmetry of the system around the mid-plane.

We discuss in the next section in detail how to use $\theta(x,z)$ to 
calculate non-linear optical effects. However, we note that in general 
the larger the values of $\theta(x,z)$ measured in some sense the larger 
will be the optical effect. It is therefore of some interest to monitor the 
behavior of $\theta(x,z)$ as a function of system parameters.

\begin{figure}[ht]
    \centering
        \includegraphics[width=0.45\textwidth]{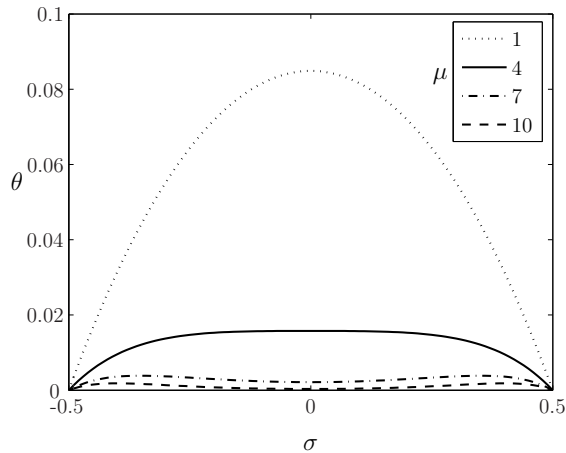}
    \caption{Functional dependence of angular distortion from homeotropic texture, in the limit of low field, for different
    values of the non-dimensionalized grating wave number $\mu$. This figure
    shows the component of the angular distortion out-of-phase with the intensity modulations.}
    \label{fig:theta_functional}
\end{figure}

We plot the z-dependence of the out-of-phase component of $\theta(x,z)$ in Fig.\ref{fig:theta_functional}. One might expect a roughly sinusoidal dependence, with a maximum  at the cell mid-plane. And indeed, for closely matched incident beams, with a low wave-number 
interference pattern, this is what occurs. But when the non-dimensional grating wave-vector $\mu$ is larger than unity, the sinusoidal dependence no longer holds. By $\mu=4$, the response
is flattened, and by $\mu \sim 6$  the profile has developed a double hump structure. Not only is the shape unexpected, but the magnitude is reduced in this regime, and as discussed in the last paragraph, this should  (and, as we shall see below,  does) lead to a reduced non-linear optical effects for larger $\mu$.

 From eq.(\ref{theta2}) we observe that  $\ds \Theta~=~ \left(\frac{1}{\cosh (\mu/2)} - \frac{1}{\cosh (k/2)}\right)$ can be regarded a figure of merit for the degree of distortion of the liquid  crystal. This is a measure of the amplitude of the response at the mid-plane of the cell. We note that $\Theta$ technically only measures the 
 out-of-phase distortion, and furthermore even then the magnitude of the distortion is not maximal at the mid-plane of the cell. Nevertheless it serves in a rough and ready way as a surrogate for
 the magnitude of the grating response to optical probes.
\begin{figure}[ht]
    \centering
        \includegraphics[width=0.45\textwidth]{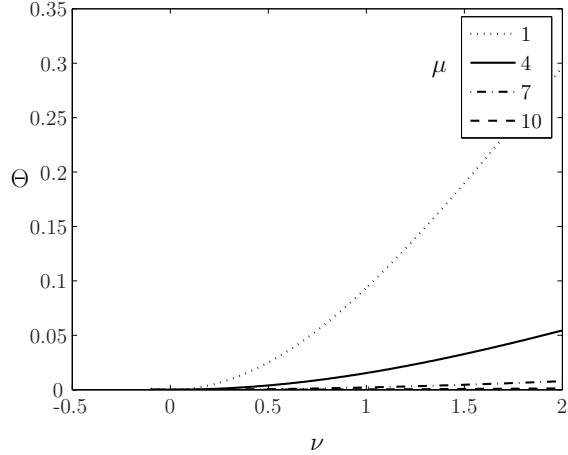}
    \caption{Dependence of the amplitude of the component of the angular distortion out-of-phase with the optical intensity,
    as a function of non-dimensionalized voltage}
    \label{fig:theta_nu}
\end{figure} In Fig. \ref{fig:theta_nu} we plot $\Theta$ as a function of voltage.
%It is interesting to note that $\Theta$ is not monotonic as a function of voltage.

\section{Diffraction of light beams}

\subsection{Formulation of problem}
We now consider light beam propagation of each of the two waves  through the
(now) weakly non-uniform anisotropic liquid crystal cell. We suppose the
wave incident from the vacuum to have  wave number $\ds
\mathbf{k}= k\hat{\mathbf{k}}$, with $k=2\pi/\lambda=\omega/c$,
where the angular frequency $\omega$ and the speed of light $c$
take their usual meanings. In the presence of a uniform liquid
crystal, the light will be refracted into an ordinary
(\textit{o}-) and an extraordinary (\textit{e}-) wave. The
ordinary wave is polarized perpendicularly to the plane of
incidence (in the $y$ direction) and the refractive index which
corresponds to this wave is $ n_o = \sqrt{\varepsilon_{\perp}}$.
For the extraordinary wave the effective refractive index is given
by:
\begin{equation}
    n_{eff}(\psi) = \frac{\sqrt{\varepsilon _{\vert \vert } \varepsilon _\bot }}{\sqrt{\varepsilon
_{\vert \vert } \cos ^2\psi +\varepsilon _\bot \sin ^2\psi }},
\label{refractive_index}
\end{equation}
where $\psi$ is the angle between the director and the direction
of propagation inside the medium.

The effect of the non-uniformity will be to modulate the amplitude
of the wave in the plane of the outgoing surface, and hence in
addition to the refraction,  diffraction will also occur. The
purpose of this section is to calculate the magnitude of this
diffraction. There are in fact two waves,  but first we 
discuss the effect of the modified director on each individual wave.

\subsection{The dielectric function}
The dielectric function is given by
\be
\varepsilon _{ij} =\varepsilon _\bot \delta _{ij} +\varepsilon _a
n_i n_j \label{dielectrictensor1},\ee with $\varepsilon _a
=\varepsilon _{\vert \vert } -\varepsilon _\bot $ and director
components $n_i$.  In the limit of interest in this paper, the
director deviations from the initial homeotropic alignment are
small. Then
\be
n = \left[ {\sin \theta \left( {x,z} \right),0,\cos \theta \left(
{x,z} \right)} \right] \approx \left[ {\theta \left( {x,z}
\right),0,1} \right]. \label{director1} \ee The dielectric
function now simplifies to \be \label{dielectric tensor2}
\hat{\e}= \hat{\e}_0 +  \theta(x,z) \hat{\e}_1, \ee or
alternatively:
\be
\hat{\e} = \left(
\begin{array}{ccc} \e_{\perp} &0&0\\ 0&\e_{\perp}&0\\
0&0&\e_{||}\end{array}\right) + \theta(x,z)\left(
\begin{array}{ccc} 0 &0&\e_{a}\\ 0&0&0\\
\e_{a}&0&0\end{array}\right). \label{director2} \ee

\subsection{Geometrical Optics}
The theoretical strategy  involves determining perturbations
around the transmission through the pure homeotropic (i.e.
$\hat{\e}_0$) system. The characteristic length for director
inhomogeneity in the $z$-direction is the cell thickness $L$. In
the $x$-direction the corresponding characteristic length is the
grating period $\Lambda =2\pi /q$. We shall use the Geometrical
Optics Approximation (GOA) \cite{panasyuk03, kraan, kravtsov}, valid in the limits
$\lambda <<L$ and $\lambda <<\Lambda $.

We seek solutions to the Maxwell equations \be \label{maxwell}
\nabla \times \mathbf E=-\mu \frac{\partial \mathbf H}{\partial t}
;\;\;\;\;
 \nabla \times \mathbf H=\frac{\partial }{\partial t}\varepsilon
_0 \hat {\varepsilon }\mathbf E \ee in the following forms:
\begin{eqnarray}
	 \mathbf  E\left( {\mathbf{r},t} \right)&=&\exp (
	-i\omega t+ i k S( \mathbf{r} )) \mathbf E_0;\nonumber \\
	 \mathbf H\left( {\mathbf{r},t} \right)&=&\exp ( -i\omega t+ i k S(
	\mathbf{r} )) \mathbf H_0
	\label{eikonalform}
\end{eqnarray}
 The term $S( \mathbf{r} )$ is the optical path length or \textit{eikonal},
 and the local direction of the wave vector is given by $\nabla S( \mathbf{r} )$.

 %We include in the solution term $ik_0 f\left( \mathbf{r}\right)$ that means a small phase change due to modulation of refractive index of liquid crystal. Then $\nabla f$ is the vector of gradient of phase and it is directed in the wave propagation direction.
  Substituting eqs.(\ref{eikonalform})  into the Maxwell equations (\ref{maxwell}), we
  obtain the following pair of equations:
\be
k \nabla S( \mathbf{r} )\times \mathbf E_{0} =\mu _0 \omega
\mathbf H_{0};\;k \nabla S( \mathbf{r} )\times
 \mathbf H_{0} =-\varepsilon _0 \hat {\varepsilon }\omega \mathbf
 E_{0}.
\ee The magnetic field $\mathbf H_0$ can now be eliminated,
yielding a homogeneous equation for $\mathbf E_0$:
\be
\nabla S( \mathbf{r} ) \times \left( {\nabla S( \mathbf{r} )
\times \mathbf E_0 } \right)+\hat {\varepsilon }\mathbf E_0 =0.
\label{eq:wave} \ee
%where we defined vector: $\mathbf p=\left( {\dfrac{c}{\omega }\mathbf{k}+\nabla f} \right)$.

Eq.(\ref{eq:wave}) is a homogeneous system of linear equations for
the electric field components, analogous to a vector Helmholtz
equation. In general solutions to this equation will be trivial
and uninteresting. However, there are non-trivial solutions,
corresponding to optical traveling waves,  if the determinant ofling waves,  if the determinant of
this set of equations is null.

In fact the determinant factorizes. An $E$ eigenvector in the $y$
direction corresponds to the ordinary wave. The perturbations in
the dielectric tensor do not affect transmission of the ordinary
wave through the sample, and we shall not be interested in this
mode of transmission. The $E$ eigenvector in the $x-z$ plane (i.e.
the plane of incidence) corresponds to the extraordinary
(\textit{e}-) wave. The pair of homogeneous equations are:

\be
\label{eq:wave2} \left[\begin{array}{cc} \e_{\perp}-(\partial_z S
)^2& \partial_x S\partial_z S +\e_a \theta(x,z)
\\ & \\ \partial_x S\partial_z S+\e_a \theta(x,z)&\e_{||}-(\partial_x S)^2
\end{array}\right]\left[\begin{array}{c} E_x \\ \\ E_z
\end{array}\right]=0
\ee The null determinant condition appropriate to the \textit{e}-
wave now reduces to\footnote{This equation is a derivative of what
in the literature is usually known as the eikonal equation.}:
\be
 {\left( {\varepsilon _\bot -(\partial_z S)^2 }
\right)\left( {\varepsilon _{\vert \vert } -(\partial_x S)^2 }
\right)-\left( {\partial_x S \partial_z S +\varepsilon _a
\theta(x,z) } \right)^2} =0. \label{eq:eikonal} \ee

\subsection{Perturbation Theory}
In the absence of the director modulation, the \textit{e}-wave is
directly transmitted. We consider eq.(\ref{eq:eikonal}) as  a
perturbation of this process. We therefore recast
eq.(\ref{eq:eikonal}) to  lowest order in $\theta(x,z) $:
\be
\frac{(\partial_x S)^2 }{\varepsilon _{\vert \vert }
}+\frac{(\partial_z S)^2 }{\varepsilon _\bot }+2\left(
\frac{\varepsilon _a }{\varepsilon _{\vert \vert } \varepsilon
_\bot}\right)\left(\partial_x S\;
\partial_z S\right)\theta(x,z)=1 \label{eq:e_wave} \ee

In the spirit of the WKB approximation, the solution of
eq.(\ref{eq:e_wave}) can be expressed as the sum of an unperturbed
\textit{e}-wave, plus a small phase change $f(\mathbf{r})$ which
can be ascribed solely to the modulation. Thus:
\be
S(r)= S_0(r) + f(\mathbf{r}), \label{eq:eikonal2} \ee where $ S_0$
obeys the equation \be \frac{(\partial_x S_0)^2 }{\varepsilon
_{\vert \vert } }+\frac{(\partial_z S_0)^2 }{\varepsilon _\bot }=1
\label{eq:e_wave0} \ee

We note also that the effective refractive index $n_{eff}$ can be
defined as follows: \be \label{neff}(\nabla S_0)^2 = n^2_{eff}.
\ee Combining eqs.(\ref{eq:e_wave0}) and (\ref{neff}) yields the
well-known expression for the refractive index
(\ref{refractive_index}). The wave vector inside the medium is now
given by \be \label{mediumwavevector} \mathbf{k'} =k\nabla S_0=
k[\partial_x S_0,0,
\partial_z S_0]= k' [\sin\psi,0, \cos \psi], \ee
with $k'=kn_{eff}$.

Combining eqs.(\ref{eq:e_wave}),(\ref{eq:eikonal2}) and
(\ref{eq:e_wave0}), we can obtain the leading order equation for
the phase $f(x,z)$:
\begin{eqnarray}
\label{eq:extraphase1}
\left(\partial_xS_0\right)\frac{\partial_xf}{\varepsilon _{||} }&+&\left(\partial_zS_0\right)\frac{\partial_zf}{\varepsilon _\bot }
 =\nonumber \\
&-&\left(\partial_xS_0\right)\left(\partial_zS_0\right)
\frac{\varepsilon _a \theta \left( {x,z} \right)}{\varepsilon
_{||} \varepsilon _\bot }.
\end{eqnarray}

We now substitute eq.(\ref{mediumwavevector}) into
eq.(\ref{eq:extraphase1}), yielding:
\bea
\left( \frac{\sin\psi}{\varepsilon _{||} }\right)\frac{\partial
f}{\partial x}&+&\left(
 \frac{\cos\psi}{\varepsilon _\bot }\right)\frac{\partial f}{\partial z}
 =\nonumber \\
 &-& \sin\psi\cos\psi
\left(\frac{\varepsilon _a n_{eff}}{\varepsilon _{||} \varepsilon
_\bot }\right)\theta \left( {x,z} \right). \label{eq:extraphase2}
\eea

Eq. (\ref{eq:extraphase2}) can be solved using the method of
characteristics. The left hand side of this equation can be
transformed into a total derivative: 
\bea \label{totalderiv1}
\frac{df}{dz}&=&\frac{\partial f}{\partial z}+
\left(\frac{dx}{dz}\right)\left(\frac{\partial f}{\partial
x}\right)=\nonumber \\  
\frac{\partial f}{\partial z}&+& 
\left(\frac{\e_{\perp}\sin\psi}{\e_{||}\cos\psi}\right)\left(\frac{\partial
f}{\partial x}\right)=\frac{\partial f}{\partial z}+
\tan\psi'\left(\frac{\partial f}{\partial x}\right). \eea Combining
eqs.(\ref{eq:extraphase2}) and (\ref{totalderiv1}) yields:
\be
\label{totalderiv2} \frac{df}{dz}=-\sin\psi
\left(\frac{\e_a}{\e_{||}}\right)n_{eff}\theta(x,z), \ee where now
the quantities $x$ and $z$ in this equation are explicitly related
by
\be
\label{poyntingpath} \ds \frac{dx}{dz}=\tan\psi', \ee defined in
eq.(\ref{totalderiv1}). Eq.(\ref{poyntingpath}) allows a family of
solutions
\be
\label{poyntingpath1} x(z,x_0)=x_0+z\tan\psi'. \ee We note that
each member of this family of solutions represents a wave entering
the liquid crystal sample at position $x_0$. The direction of
\textit{wave propagation} is given by the angle $\psi$. However,
because this medium is anisotropic, the angle of the
\textit{energy propagation}, given by the Poynting Vector, is
determined by the angle $\psi'$. The family of solutions
$x\left(z,x_0\right)$ corresponds to paths in the undistorted
anisotropic medium with different $x_0$ travelling in the
direction of the Poynting Vector.

Now we can solve eq. (\ref{totalderiv2}) directly, by integrating
the right hand side, yielding:
\be
\label{phase_retardation1}
 f\left( {x,z} \right)=-\sin\psi\left(\frac{\varepsilon _a n_{eff} }{\varepsilon _{\vert
 \vert } }\right)\int_0^z {\theta \left( {x'(z',x_0),z'} \right)} dz',
 \ee
where the integration path is such that
$x=x_0+z\tan\psi';\;\;x'=x_0+z'\tan\psi'$, and hence
\be
\label{poyntingpath2} x'=x-(z-z')\tan\psi'. \ee
Thus
\bea
\label{phase_retardation2}
 f\left( {x,z} \right)&=&-\sin\psi\left(\frac{\varepsilon _a n_{eff} }{\varepsilon _{\vert
 \vert } }\right)\nonumber \\
 &&\times\int_0^z {\theta \left( {x-(z-z')\tan\psi',z'} \right)} dz',
\eea

The key quantity of interest is the phase retardation of the beam
as it leaves the  cell, i.e. at  $z=L$.  We now rewrite
eq.(\ref{phase_retardation1}), so as to express this quantity
directly:
\bea
\label{phase_retardation3} f\left( {x,L}
\right)&=&-\sin\psi\left(\frac{\varepsilon _a n_{eff} }{\varepsilon
_{\vert
 \vert } }\right)\nonumber \\
 &&\times\int\limits_0^L \theta \left(x-(L-z')\tan\psi',z' \right) dz' \eea

\subsection{Diffraction Pattern}
The formula (\ref{phase_retardation3}) applies to all incident light beams. We confine our interest to the cases
in which there are two incident beams, with wave numbers $k_1$ and $k_2$, with $k_{1x}-k_{2x}=q$.
Eq.(\ref{phase_retardation3}) permits the calculation the light fields of the beams as they exit the liquid crystal cell.
By suitably decomposing these  light field into Fourier components, it is possible to identify the amplitudes of particular diffracted beams.

The light field along the plane $z=L$ is modulated  by the factor
\begin{eqnarray} \label{exp_eikonal}
 \exp[ i kS(x,L)]=\exp[ i (\mathbf{k}'\cdot\mathbf{r}  + kf(\mathbf{r}))]=\nonumber \\ 
 \exp[i(k_x x+k'_z L + kf(x,L))]= \exp[(i(k_x x +\delta \phi_0 + \delta \phi_1)],
\end{eqnarray}
 where $\delta \phi_0 = k'_z L = k n_{eff} L \cos \psi $, and $\delta \phi_1 = kf(x,L) $;
\be
\delta \phi _0 =\frac{\sqrt {\varepsilon _{\vert \vert
} \varepsilon _\bot } }{\left( {\varepsilon _{\vert \vert } \cos ^2\psi
+\varepsilon _\bot \sin ^2\psi } \right)^{1/2}}\left( {kL\cos \psi }
\right)
\label{delta_phi0}
\ee
and

\bea
 \delta \phi _1 &=&- k\frac{\sqrt {\varepsilon _\bot }
( {\varepsilon _{\vert \vert } -\varepsilon _\bot })\sin \psi
}{\sqrt {\varepsilon _{\vert \vert } } ( {\varepsilon _{\vert
\vert } \cos ^2\psi +\varepsilon _\bot \sin ^2\psi }
)^{1/2}}\nonumber \\
&&\times\left[\int\limits_0^L \theta \left(x-(L-z')\tan\psi',z'
\right) dz'  \right]. \label{delta_phi1} \eea 
The quantity $\delta \phi _1(x)$ is the varying component of the additional phase of the incident beam, following from the director modulation in the liquid crystal medium. We now  recall that from eq.(\ref{theta2}) the director profile can be written in a sinusoidal form: $\theta(x,z) \propto \cos(qx + \Delta)$. Combining this result with eq.(\ref{delta_phi1}) yields the
result:
\be
\label{delta_phi2}
 \delta\phi_1(x) = B \cos(qx + \tilde{\Delta}),
\ee 
where the phase modulation parameters $B>0$ and $\tilde{\Delta}$ are respectively the amplitude and phase of the additional phase $\delta \phi _1(x)$:
\be
\label{definitions}
B = \sqrt{A^2+C^2};\;\;\;  \ds \tan \tilde{\Delta} =
-\frac{A}{C}.\ee After complicated but straightforward algebra using eqs. (\ref{theta2}), (\ref{delta_phi1}), expressions for the quantities $A$ and $C$ can be derived:
\begin{widetext}
\begin{equation} \ds
    A =  -kL\frac{\Phi_1}{\Phi_0}\frac{\mu
( {\varepsilon _{\vert \vert } -\varepsilon _\bot })\sin \psi
}{( {\varepsilon _{\vert
\vert } \cos ^2\psi +\varepsilon _\bot \sin ^2\psi }
)^{1/2}}\int_{-1/2}^{1/2}\left[{\begin{array}{l} \cos \frac{\delta
    }{2}\left\{
{\frac{\cosh \mu \sigma }{\cosh \mu /2}-\frac{\cosh \kappa \sigma }{\cosh
\kappa /2}} \right\}\cos \left(\mu (\sigma-\frac12) \tan\psi' \right) \\ \quad \quad \quad \quad \quad \quad  \quad \quad \quad -\\ \sin \frac{\delta }{2}\left\{
{\frac{\sinh \mu \sigma }{\sinh \mu /2}-\frac{\sinh \kappa \sigma }{\sinh
\kappa /2}} \right\}\sin \left(\mu (\sigma-\frac12) \tan\psi' \right)   \end{array}} \right] d \sigma
\label{eq:A}
\end{equation}
\begin{equation}
    C =  -kL\frac{\Phi_1}{\Phi_0}\frac{\mu
( {\varepsilon _{\vert \vert } -\varepsilon _\bot })\sin \psi
}{ ( {\varepsilon _{\vert
\vert } \cos ^2\psi +\varepsilon _\bot \sin ^2\psi }
)^{1/2}}\int_{-1/2}^{1/2}\left[{\begin{array}{l} \cos \frac{\delta }{2}\left\{
{\frac{\cosh \mu \sigma }{\cosh \mu /2}-\frac{\cosh \kappa \sigma }{\cosh
\kappa /2}} \right\}\sin \left(\mu (\sigma-\frac12) \tan\psi' \right) \\ \quad \quad \quad \quad \quad \quad  \quad \quad \quad +\\ \sin \frac{\delta }{2}\left\{
{\frac{\sinh \mu \sigma }{\sinh \mu /2}-\frac{\sinh \kappa \sigma }{\sinh
\kappa /2}} \right\}\cos \left(\mu (\sigma-\frac12) \tan\psi' \right)   \end{array}} \right] d \sigma
\label{eq:C}
\end{equation}
\end{widetext}
Now eq.(\ref{exp_eikonal}) can be rewritten, using eq.(\ref{delta_phi2}), so as explicitly identify different components of
the diffracted wave. The key relation is the Jacobi-Anger
expansion \cite{Abramowitz}: \be \label{Jacobi_Anger}
  \exp \left( {iz\cos \phi } \right)=\sum\limits_{n=-\infty }^{n=+\infty } {\left( i \right)^nJ_n \left( z \right)\exp \left( {in\phi } \right)}
\ee
 Combining eq.(\ref{Jacobi_Anger}) with eqs.(\ref{exp_eikonal}), (\ref{delta_phi0}), (\ref{delta_phi1}) and (\ref{delta_phi2}) yields the expansion for the electric field at the output surface:
\begin{eqnarray}
    E_{out}  = E_0\exp(i \delta\phi_0+i k_x x) \exp(i \delta\phi_1) =\nonumber \\
E_0\exp(i \delta\phi_0+i k_x x)\!\!\!\!\sum_{n=-\infty}^{\infty}(i)^n\!\! J_n(B) \exp(in q x +in \tilde{\Delta})
    \label{eq:el_field}
\end{eqnarray}
 Now we can identify \cite{born80} terms in this expansion with the amplitudes $X^{(n)}$ and phases $\delta^{(n)}$ of outgoing waves in
 the diffraction pattern:
\be
 X^{(n)} = (i)^n J_n(B)
 \label{outgoingamplitude}
\ee
 and
\be
 \delta^{(n)} =  \delta\phi_0 +n \tilde{\Delta}.
 \label{outgoingphase}
\ee
 The electric field in the diffracted wave of order $n$  then takes the form:
\be
\label{diffraction1} \frac{E_n}{E_0}= X^{(n)} \exp
i\left(\mathbf{k}^{(n)}\cdot\mathbf{r}+\delta^{(n)}\right), \ee
where $\mathbf{k}^{(n)}$ is the wave number of the diffracted wave
of order $n$, with $k^{(n)}_{x}=k_x+nq$, and
$|\mathbf{k}^{(n)}|=|\mathbf{k}|$.

\subsection{Beam Coupling}

We now return to the original problem (\ref{B1}) in which there
are two incident waves with wave numbers ${\rm {\bf k}}_1,{\rm
{\bf k}}_2 $, and with $k_{1x}-k_{2x}=q$. Beam coupling
corresponds the diffraction of waves from incident wave
$\mathbf{k}_1$ to outgoing wave
$\mathbf{k}_2=\mathbf{k}_1-q\mathbf{e}_x$, and from incident wave
$\mathbf{k}_2$ to outgoing wave
$\mathbf{k}_1=\mathbf{k}_2+q\mathbf{e}_x$. Thus the diffracted
wave of order $-1$ from $\mathbf{k}_1$ adds coherently with the
directly transmitted wave $\mathbf{k}_2$, and  the diffracted wave
of order $+1$ from $\mathbf{k}_2$ adds coherently with the
directly transmitted wave $\mathbf{k}_1$. Equivalently, using the
notation of the last section,  $\mathbf{k}_1=\mathbf{k}^{(+1)}_2$
and $\mathbf{k}_2=\mathbf{k}^{(-1)}_1$.

We are thus able to use  terms from the diffraction expression
eq.(\ref{diffraction1}) to evaluate the amplitudes of the outgoing
waves directions $\mathbf{k}_1$ and $\mathbf{k}_2$. We find:
\begin{eqnarray}
\label{intensity1} E_{out}(\mathbf{k}_1)=E_{01} X^{(0)}
\exp(i\delta^{(0)}) + E_{02} X^{(+1)} \exp(i\delta^{(+1)})
 \nonumber \\
 \ \\ E_{out}(\mathbf{k}_2)=E_{01} X^{(-1)}
\exp(i\delta^{(-1)}) + E_{02} X^{(0)} \exp(i\delta^{(0)})
\nonumber
\end{eqnarray}

We note that in principle the quantities $X^{(n)}$ depend on the
value of the incident wave number. However, in the two-beam
coupling case discussed here, the incident waves have wave numbers
very nearly equal to each other, and so we may consider the
quantities $X^{(n)}$ to be the same for each incident wave.

From eq.(\ref{intensity1}), we can evaluate the outgoing wave
intensities, using the relation $I = \mathbf{E} \cdot \mathbf{
E}^*$. Using eq.(\ref{outgoingamplitude}), We find:
\begin{multline}
    I_{out}(\mathbf{k}_1) = (E_{01}\exp(i \delta^{(0)})J_0(B) + i E_{02}\exp(i \delta^{(+1)}) J_1(B)) \\((E_{01}^*\exp(-i \delta^{(0)})J_0(B) - i E_{02}^*\exp(-i \delta^{(+1)})J_1(B))),
\end{multline}
or \bea \label{Iout}
    I_{out}(\mathbf{k}_1)\!\!\! &=&\!\!\! I_1J_0^2(B)+I_2J_1^2(B) - 2\sqrt{I_1 I_2} J_0(B) J_1(B)
    \sin( \tilde{\Delta}); \nonumber \\
&& \\
     I_{out}(\mathbf{k}_2)\!\!\! &=&\!\!\! I_2J_0^2(B)+I_1J_1^2(B) +
     2\sqrt{I_1 I_2} J_0(B) J_1(B) \sin( \tilde{\Delta}),\nonumber
\eea
where $\tilde{\Delta}$ is as defined in eq.(\ref{definitions}).

It is usual to consider one of the beams as the \textit{pump} beam
and the other as the \textit{probe} beam. Without loss of
generality, we shall suppose that $\mathbf{k}_1$  corresponds to
the pump beam and $\mathbf{k}_2$ to the probe beam. In line with
the literature, we define  \be \label{pumpprobe}m  =
\dfrac{I_{probe}}{I_{pump}}= \dfrac{I_2}{I_1}.\ee We can now
rewrite the formulas for the outgoing beam intensities in terms of
the quantity $m$ as follows:

\begin{eqnarray}
    I_{out}(\mathbf{k}_1)\!\!\! &=&\!\!\!    I_1\left(J_0^2(B)+mJ_1^2(B) - 2\sqrt{m} J_0(B) J_1(B) \sin( \tilde{\Delta})\right); \nonumber \\ &&\label{eq:intensities}\\
      I_{out}(\mathbf{k}_2)\!\!\! &=&\!\!\! I_1\left(m J_0^2(B)+J_1^2(B) + 2\sqrt{m} J_0(B) J_1(B) \sin( \tilde{\Delta})\right).
\nonumber
\end{eqnarray}

The degree of beam coupling can now be characterized by the \textit{Gain} $g$. This is the ratio of the
intensity of the outgoing  beam in the direction of the probe beam in the presence of the pump beam to the intensity  of the same beam in the absence of the pump beam. In the context of this paper, in which we do not consider reflection and refraction at the cell walls, the quantity $g$ is defined as:
\be
\label{definition2}
g=\frac{I_{out}(\mathbf{k}_2)}{I_2} = J_0^2(B)+\frac{1}{m}J_1^2(B) + 2\frac{1}{\sqrt{m}} J_0(B) J_1(B) \sin(\tilde{\Delta}).
\ee

A related quantity is the \textit{Diffraction Efficiency} $\eta$, which measures the strength with which the grating  diffracts   the probe beam. The formal definition is the ratio of the intensity of the diffracted probe beam (i.e. in the direction of the pump beam) to that of the incoming  probe beam.  From eqs.(\ref{outgoingamplitude}) and (\ref{diffraction1}), this is:
\begin{equation}
 \eta = |X^{(-1)}|^2 = J_1^2(B).
 \label{de}
\end{equation}

For ease of presentation of our results, it is also convenient to define quantities $\eta'$ and $g'$. These quantities are respectively analogous to $\eta$ and $g$, but with the roles of the pump and probe beams exchanged.

\section{Results}

\subsection{Analytical study of the behavior of intensities. }
 First we discuss diffraction from a single beam. The transmitted energy is divided between beams of different orders $n$. The amplitude of diffracted beams is given by $|X^{(n)}|^2 \propto J_n^2(B)$ (eq.(\ref{outgoingamplitude})), where we recall, from eq.(\ref{definitions}) that $B$ is the amplitude of the additional phase variations induced by the spatial modulations in the liquid crystal layer.
 We introduce the \textit{energy conservation parameter} 
 \begin{equation}
	s_n(B) = \sum_{k=-n}^{k=n} J_k^2(B).
\end{equation}
 The quantity $s_n(B)$ describes the proportion of transmitted energy that is distributed between diffracted beams of orders from $-n$ to $n$. 
  We note that in the limit $n\rightarrow \infty$ necessarily $s_n \rightarrow 1$. 
  
\begin{figure}[ht]
	\centering
		\includegraphics[width=0.45\textwidth]{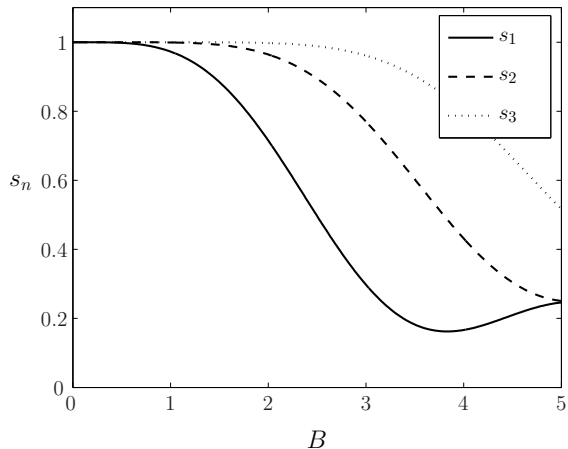}
	\caption{The dependence of energy conservation parameter $s_n(B)$ on phase modulation parameter $B$. Solid line: $s_1$, dashed line $s_2$ and dotted line $s_3$. See text for further discussion.}
	\label{fig:s1s2s3}
\end{figure}

  In Fig.\ref{fig:s1s2s3} we show the dependence of $s_n$ on the phase modulation parameter $B$ for low $n$. For small $B$ ($B\leq 1 $),  $s_1(B)\approx 1$. In this regime almost all  transmitted energy is either in  the directly transmitted beam, or in the first-order diffracted beam. This is  the regime in which energy exchange between two beams is most effective. 
  
  As $B$ increases, an increasing proportion of the transmitted energy is transfered to outlying diffracted beams. The quantity $s_2$ remains essentially unity until $B\approx 1.5$, while $s_3$ only noticeably departs from unity at $B \approx 2.5$. We shall return to the problem of the asymptotic behavior of $s_n(B)$ for large $n, B$ elsewhere. For this study, however, we shall be interested in the small $B$ regime.

 We now analyze the effect of energy exchange in the presence of both incident beams. We shall take equal intensities in eq.(\ref{eq:intensities}) for incident beams $m = 1$. 
 In principle the parameter $m$ [eq.(\ref{pumpprobe})], measuring the ratio of the intensities of the beams, can take any value. In our calculations we shall suppose $m=1$; this corresponds to equal intensity beams. This will enable us to make contact with previous studies \cite{N2}, which have also used this value. In addition it is easy to monitor energy transfer between beams. We note that in devices we may well expect that $m \ll 1$, so that a large reservoir of pump beam energy is available to amplify a given probe beam.
  
 The degree of energy transfer is critically dependent on the quantity $\tilde{\Delta}$, defined in eq.(\ref{definitions}). When $\tilde{\Delta}=0$, the modulation of the phase retardation is in-phase with the intensity modulation due to the beam interference. If $m=1$, eq.(\ref{eq:intensities}) implies that there is no net energy exchange between the beams. This is consistent with general intuition \cite{Yeh} that a phase difference between the intensity and dielectric modulations is required for beam coupling. Interestingly, although we do not pursue this here, this rule no longer holds for the $m \neq 1$ case.
  
 The maximum energy transfer between beams occurs when $\sin(\tilde{\Delta}) = 1$. In this case the two outgoing beams obey the following rule:
\begin{eqnarray}
    I_1 &\propto& \left(J_0(B) -  J_1(B)\right)^2 \nonumber \\
    I_2 &\propto& \left(J_0(B) +  J_1(B)\right)^2. 
    \label{eq:max_i}
\end{eqnarray}

\begin{figure}[!ht]
    \centering
        \includegraphics[width=0.45\textwidth]{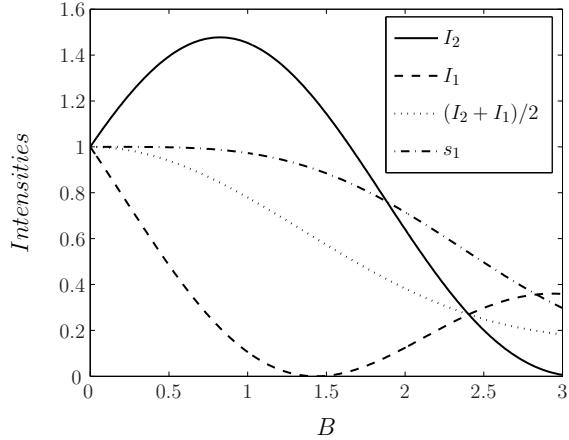}
    \caption{The behavior of functions $I_2(B)$(solid curve) and $I_1(B)$(dot curve) when $\sin(\tilde\Delta) = 1$.}
    \label{fig:I1I2}
\end{figure}

    The behavior of these functions is shown in Fig.\ref{fig:I1I2}. The function $I_2(B)$ (solid curve on the graph) reaches a maximum value of  $\approx 1.48$ at $B \approx 0.85$. We also plot $I_1(B)$, and note that this reaches a minimum(at zero) for $B \approx 1.4$. The quantity 
\begin{equation}
 \overline{I}(B)=	\frac12 (I_1(B)+I_2(B)) 
 \label{eq:I_bar}
\end{equation}
 denotes that proportion of the energy of the incident beams which remains in the two initial beams directions.
 The quantity $\overline{I}(B)$ is unity for $B=0$ (at which  there is, however, no energy exchange) and reduces steadily with B. Close to the maximum $I_2(B)\approx 1.48$ at $B \approx 0.85$,  $\overline{I}(B) \approx 0.8$, reducing monotonically to $\overline{I}(B\approx2) \approx 0.4$. However, for $B\le1$, $s_1(B)$ is essentially unity. The energy lost from the primary beams reappears as other $|n|=1$ diffracted beams. Subsequent maxima of $I_2(B)$ take values less than unity, in regimes in which a substantial proportion of the transmitted energy is lost in outlying diffracted beams. Thus the appearance of energy transfer between beams is restricted to the first maximum of $I_2(B)$.

%%%%%%%%%%%%%%%%%%%%%%%%%%%%%%%%%%%%%%%%%%%%%%%%%%%%%%%%%%%  
%\clearpage

\subsection{Dependence on external parameters}
   
   We now examine quantitatively the energy exchange process, using experimentally plausible parameters. A list of parameters in the problem is given in Table \ref{tab:parameters}, together with meanings of these parameters, and where appropriate, the numerical values that we have used in  our calculations.

\begin{table*}
	\caption{\label{tab:parameters}Table of parameters}
	\begin{ruledtabular}
	\begin{tabular}{c c p{12cm}}
	%\begin{tabular}{c c c}
	\textbf{Parameter} & \textbf{Value} &  \multicolumn{1}{c}{\textbf{Description}}    \\ \hline
	\hline
	 $\lambda$ & $0.63\mu m$ & Wavelength of incident beams   \\ \hline
	 $L$       & $20\mu m $ & Thickness of the film \\ \hline
	 $\varepsilon_{\bot}$, $\varepsilon_{\|}$ & $1.5^2$, $1.7^2$  & Dielectric permittivities of the liquid crystal \\ \hline
	 $\psi$ & variable & Angle of propagation inside liquid crystal \\ \hline
	 $\delta$  & variable &  Phase shift between the interference patterns at top and bottom surfaces. This is the surrogate for the angle of incidence which we do not include explicitly.\\ \hline
	 $\gamma$ & variable & Half-angle between beams defining the dimensionless grating wave-vector $\mu = \tilde{q}L = 2 k L\sqrt{\frac{\varepsilon_{\bot}}{\varepsilon_{||}}} \cos\psi\sin\gamma $ \\ \hline
	 $\mu$ & variable & Non-dimensional grating wave vector $\mu =\tilde{q}L$. For  $\gamma \approx 2.4^0$, $\mu \approx 6$. \\ \hline
	 $\nu$ & 1 & Dimensionless voltage $\ds \nu =L E_0(\frac{\e_a}{K})^{1/2}$\\ \hline
	 $m$   & 1 & Ratio of intensities of incoming beams \\
	\end{tabular}
	\end{ruledtabular}
\end{table*}

 The final element in the theory enabling comparison with experiment is the response of the surface potential $\Phi_1$ to the local beam intensity. We do not however have a microscopic photoelectrochemical theory to describe this process. In principle $\Phi_1$ is a measurable quantity, although in practice the measurement may be difficult to carry out.
 
 In our initial calculations we suppose  $\Phi_1/\Phi_0=1$ (eq.\ref{theta2}) and the external voltage $\nu=1$. 
   
  We first investigate the dependence of energy exchange effect on $\psi$. We set the grating period $\mu=6$, which corresponds to a grating wavelength $\Lambda$ equal to the cell thickness $L$. To see the influence of both the in-phase and out-of phase components we choose the phase shift of the interference patterns $\delta=\pi/2$. 
  
% \item In articles about theory of photorefraction investigated gain for $m=1$. See e.g.: Gary P. Wiederrecht ``PHOTOREFRACTIVE LIQUID CRYSTALS''  (p.10); Hiroshi Ono, APL97; Simoni.   
%\end{enumerate}

\begin{figure}[ht]
    \centering
        \includegraphics[width=0.45\textwidth]{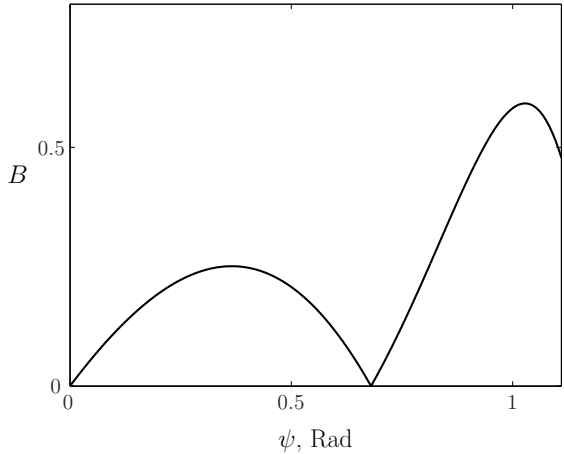}
    \caption{ Functional dependence of the phase modulation parameter $B$ as a function of the internal angle $\psi$. Fixed parameters: $\delta = \pi/2$, $\mu = 6$ and $\Phi_1/\Phi_0 = 1$.}
    \label{fig:g_psi_B}
\end{figure}

\begin{figure}[ht]
    \centering
        \includegraphics[width=0.45\textwidth]{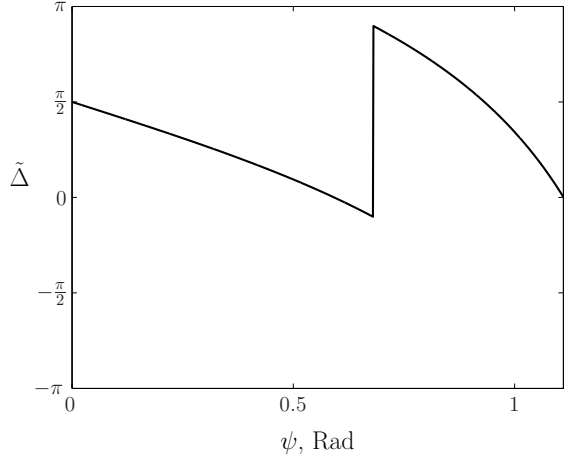}
    \caption{ Functional dependence of the phase modulation parameter $\tilde{\Delta}$ as a function of the internal angle. Other parameters are fixed: $\delta = \pi/2$, $\mu = 6$ and $\Phi_1/\Phi_0 = 1$.}
    \label{fig:g_psi_phi}
\end{figure}

 The key intermediate parameters  governing $g$ [eq.(\ref{definition2})] and $\eta$ [eq.(\ref{de})] are the phase modulation parameters $B$ and $\tilde{\Delta}$. In Fig. \ref{fig:g_psi_B} we plot the phase modulation parameter $B$ as a function of $\psi$, the angle of propagation inside the liquid crystal. The principal result from Fig. \ref{fig:g_psi_B} is that $B<1$ everywhere, which implies that only transmitted and first order diffracted beams occur. In Fig. \ref{fig:g_psi_phi} we plot $\tilde{\Delta}(\psi)$. 
 It can be seen from eq.(\ref{eq:intensities}) that the phase modulation parameter $\tilde{\Delta}$ governs the sign of the energy exchange. For $0<\tilde{\Delta}<\pi$ $\sin \tilde{\Delta} > 0$, and the probe beam is amplified by the pump beam. However for $\pi<\tilde{\Delta}<2\pi$ $\sin \tilde{\Delta} < 0$ and the pump beam is amplified. 
 In Fig. \ref{fig:g_psi_B} $B(\psi)=0$ for $\psi \approx 0.68$. At this point the quantities $A$ [eq.(\ref{eq:A})] and $C$ [eq.(\ref{eq:C})] are both equal to zero and change sign. This implies a sudden phase shift of $\pi$ in the phase modulation parameter $\tilde{\Delta}$, which indeed occurs at $\psi \approx 0.68$ in Fig. \ref{fig:g_psi_phi}.
 
\begin{figure}[ht]
    \centering
        \includegraphics[width=0.45\textwidth]{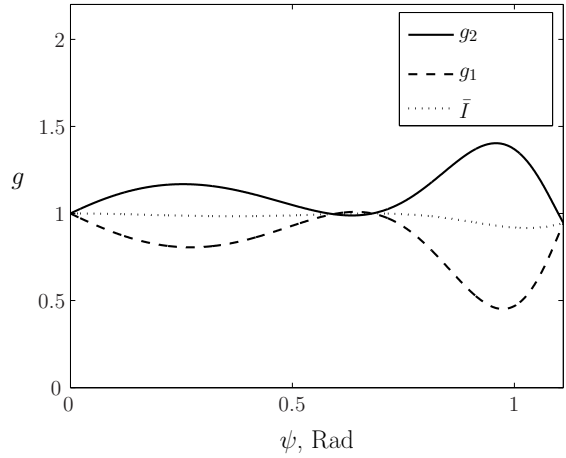}
    \caption{Energy exchange between beams as a function of the internal angle $\psi$. Solid line: gain of probe beam. Dashed line: gain of pump beam. Dotted line: $\bar{I}$ (see eq.(\ref{eq:I_bar})) represents the degree of energy conservation in the system.
     Fixed parameters: $\delta = \pi/2$, $\mu = 6$ and $\Phi_1/\Phi_0 = 1$.}
    \label{fig:g_psi}
\end{figure}

 The behavior of the gain is shown in Fig. \ref{fig:g_psi}. Maximal gain is achieved when $\sin \tilde{\Delta}(\psi) = 1$, corresponding to $\psi \approx 0.95$ (see eqs.(\ref{eq:intensities}, \ref{eq:max_i})). At $\psi = 0 $ the gain $g(\psi)=1$. We also note that at this point, from the symmetry of the system there is no energy exchange. The quantity $\bar{I}$ plotted on this graph  is almost unity everywhere. This means that there are no energy losses and all incident energy is distributed between outgoing probe beam and outgoing pump beam.

% -------------- 
  
\begin{figure}[!ht]
    \centering
        \includegraphics[width=0.45\textwidth]{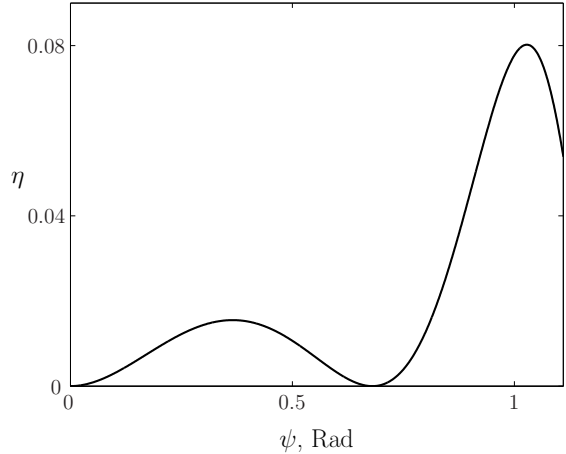}
    \caption{The dependence of diffraction efficiency $\eta$ on the internal angle $\psi$. Fixed parameters: $\delta = \pi/2$, $\mu = 6$ and $\Phi_1/\Phi_0 = 1$.}
    \label{fig:de_psi}
\end{figure}

In Fig. \ref{fig:de_psi} we plot the pump beam  diffraction efficiency. This measures the proportion of energy diffracted from the pump beam in the probe beam direction. The diffraction efficiency $\eta$ is a function  only of $B$ [eq.(\ref{de})]. 
 The maximum possible diffraction efficiency is given by $\eta(B) \approx 0.338$ at $B \approx 1.83$. This maximum value does not depend on the details of our model and remains true for thin gratings \cite{N14}. But in Fig. \ref{fig:de_psi}, the maximum value of diffraction efficiency $\eta \approx 0.08$ occurs at $\psi \approx 1.05$. The discrepancy between the theoretical maximum and this maximum can be ascribed to the fact that $B(\psi)<1$ everywhere.

\begin{figure}[ht]
    \centering
        \includegraphics[width=0.45\textwidth]{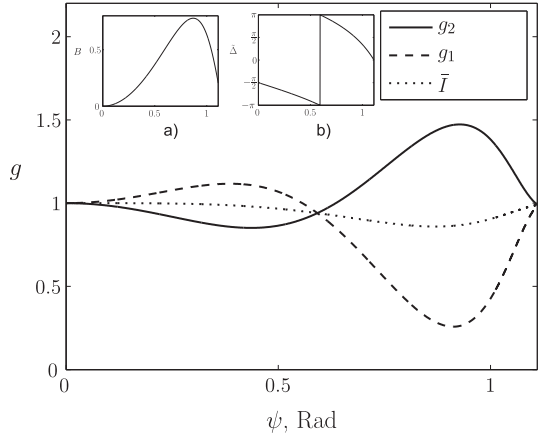}
    \caption{As for Fig. \ref{fig:g_psi}, but with $\delta =\pi$ (see text). Energy exchange between beams as a function of $\psi$. Solid line: gain of probe beam. Dashed line: gain of pump beam. Dotted line: $\bar{I}$ (see Fig. \ref{fig:g_psi}). Insets: a) Phase modulation amplitude $B$ as a function of the mean angle of propagation $\psi$; b) Phase modulation parameter $\tilde{\Delta}(\psi)$.
}
\label{fig:g_psi1}
\end{figure}

The phase shift $\delta$ [eq.(\ref{eq:delta})] also strongly affects the energy exchange characteristics. In Figs. \ref{fig:g_psi_B}-\ref{fig:de_psi} $\delta=\pi/2$ and the modulation of the dielectric and of the energy along the cell are  out-of-phase. In Fig. \ref{fig:g_psi1}, we put $\delta = \pi$; now these modulations are in-phase. Now it is possible to transfer energy from the probe beam to the pump beam, by contrast with previous case in which negative energy transfer never occurs. Thus, whereas in Fig. \ref{fig:g_psi} $g>1$ everywhere, in Fig. \ref{fig:g_psi1}, $g-1$ can take either positive or negative signs.

%
%\begin{figure}[ht]
%    \centering
%        \includegraphics[width=0.80\textwidth]{de_psi1}
%    \caption{ Figure shows the dependence of diffraction efficiency $\eta$ from the angle of propagation $\psi$.}
%\end{figure}

%\subsection{Energy exchange as a function of angle between interfering beams.}

 We now turn to the study of energy exchange as a function of the angle between interfering beams. Increasing the angle $2\gamma$ between beams increases the non-dimensional wave vector $\mu=\tilde{q}L$. Quantitatively, choosing parameters given in Table \ref{tab:parameters}, we find that for  $\gamma \approx 2.4^0$, $\mu \approx 6$.  We examine the dependence $g(\mu)$ for $\psi=0.95$, i.e. at the maximum of $g$  appropriate to Fig. \ref{fig:g_psi}.
\begin{figure}[ht]
    \centering
        \includegraphics[width=0.45\textwidth]{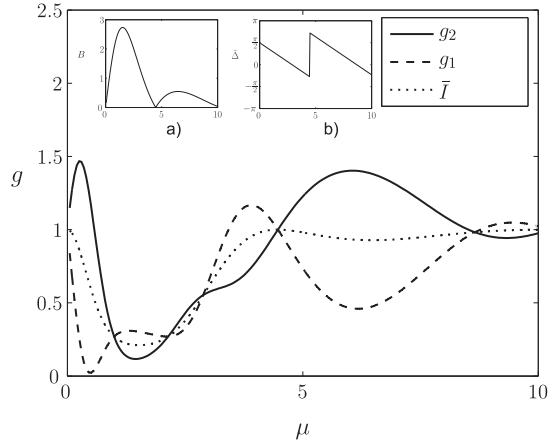}
    \caption{Energy exchange between beams as a function of $\mu$, the non-dimensional grating wave vector. Solid line: gain of probe beam. Dashed line: gain of pump beam. Dotted line: $\bar{I}$ (see Figs. \ref{fig:g_psi}, \ref{fig:g_psi1}). Insets: a) Phase modulation amplitude $B$ as a function of  $\mu$; b) Phase modulation parameter $\tilde{\Delta}(\mu)$. Fixed parameters: $\delta = \pi/2$, $\psi = 0.95$ and $\Phi_1/\Phi_0 = 1$.}
\label{fig:g_mu}
\end{figure}   
 The dependence of the gain on  $\mu$ is shown in Fig. \ref{fig:g_mu}. For  $\mu<4$, $B$ is larger than unity (see inset $a)$) and $\bar{I}<1$ (see eq.(\ref{eq:I_bar})) is less than unity. In this regime there is considerable energy loss due to probe beam diffraction into diffraction orders of order $n>1$. However for $\mu>4$, $B<1$ and the quantity $\bar{I}$ is close to one.  In this regime energy is conserved. The maximal gain is achieved at $\mu \approx 6$. This maximum is consistent  with our results from Figs. \ref{fig:g_psi_B}-\ref{fig:de_psi}. 
\begin{figure}[ht]
    \centering
        \includegraphics[width=0.45\textwidth]{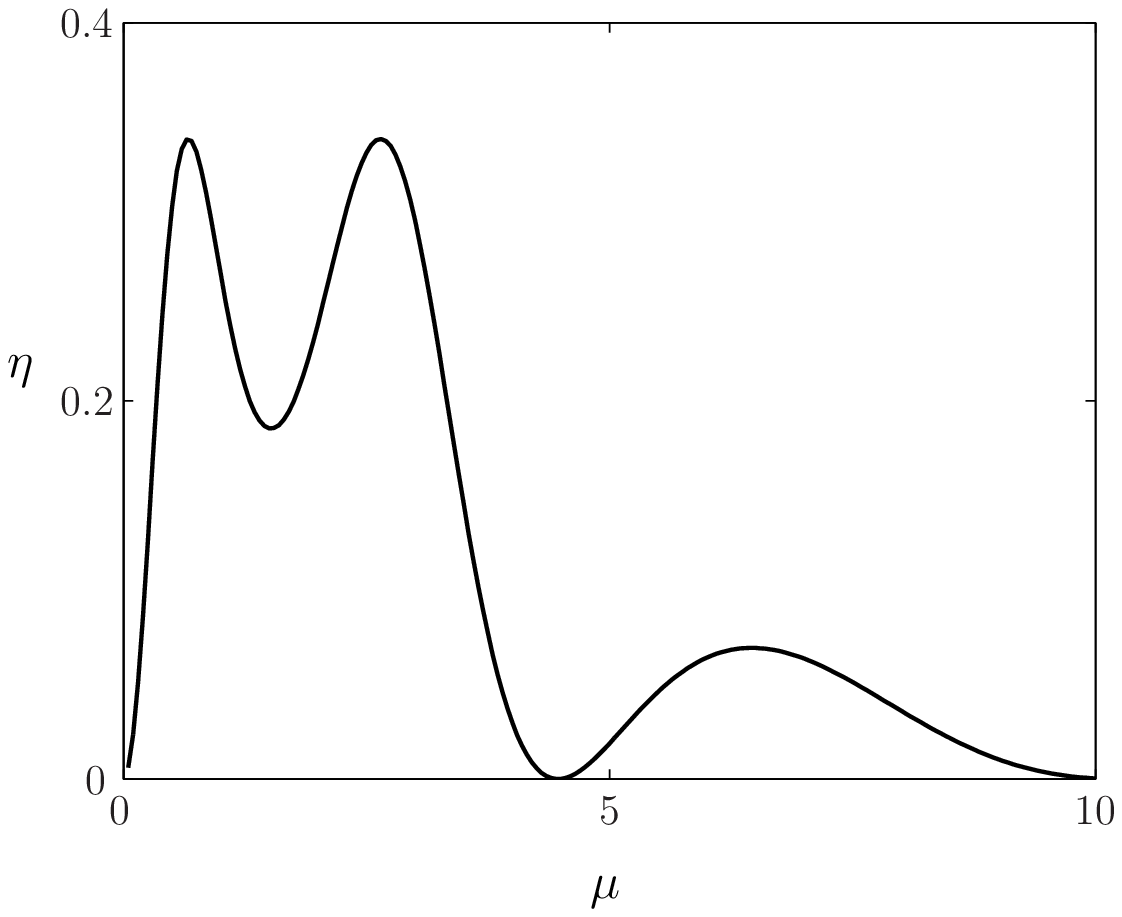}
    \caption{The dependence of diffraction efficiency $\eta$ on the non-dimensional wave vector $\mu$. Fixed parameters: $\delta = \pi/2$, $\psi = 0.95$ and $\Phi_1/\Phi_0 = 1$.}
    \label{fig:de_mu}
\end{figure}
 
   In Fig. \ref{fig:de_mu} we plot the diffraction efficiency $\eta$. The maxima of $0.338$ occur at $\mu \approx0.65$ and $\mu \approx 2.65$, corresponding to  maxima in the Raman-Nath regime \cite{raman_nath}. But in our case these maxima occur for $\mu<4$, where as we have seen above, there is considerable diffractive energy loss. The physical relevant maximum occurs for $\eta(\mu = 6.5) \approx 0.07 $. Here there is insignificant energy loss.

  The strength of the grating modulation (eq.(\ref{theta2})) depends on the ratio $\ds \frac{\Phi_1}{\Phi_0}$. In all our previous plots  (Figs. \ref{fig:g_psi_B}-\ref{fig:de_mu}) we have set the ratio $\ds \frac{\Phi_1}{\Phi_0}=1$. This ratio can be modified in two ways. In principle one can change $\Phi_1$ by changing the surface preparation. Alternatively (and more simply) one can apply an external voltage $\Phi_0$ across the liquid crystal. Here we suppose $\Phi_1$ and $\Phi_0$ to be independent quantities.  
  
  We now investigate the effect of external voltage on the energy exchange for $g(\psi = 0.95, \mu = 6)$. This corresponds to the point (see Fig.\ref{fig:g_psi}) where $g$ is maximal with respect to varying $\psi$ with other parameters as taken in Table \ref{tab:parameters}.  The optical modulation is a strong function of the director modulation. It is thus useful to analyze the director modulation as a function of external applied field.
  
   We first  make a quantitative analysis of the behavior of the director modulation as a function of voltage. The liquid crystal director distribution is given by eq.(\ref{theta2}):
\begin{equation*}
	\theta \left( {x,z} \right)=-q L \frac{\Phi _1}{\Phi_0}  \left[ {\begin{array}{l}
	 \cos \frac{\delta }{2}\sin \left( {qx+\frac{\delta }{2}} \right)\left\{
	{\frac{\cosh \mu \sigma }{\cosh \mu /2}-\frac{\cosh \kappa \sigma }{\cosh
	\kappa /2}} \right\} \\
	 \quad \quad \quad \quad \quad \quad \quad + \\
	 \sin \frac{\delta }{2}\cos \left( {qx+\frac{\delta }{2}} \right)\left\{
	{\frac{\sinh \mu \sigma }{\sinh \mu /2}-\frac{\sinh \kappa \sigma }{\sinh
	\kappa /2}} \right\} \\
	 \end{array}} \right],\eqno (27)
\end{equation*}
The voltage enters this expression explicitly through the multiplier $\ds \frac{\Phi_1}{\Phi_0}$, and implicitly through the quantity $\kappa$, where $\kappa ^2=\mu ^2+\nu ^2$, $\ds \nu =\frac{L}{\xi } = \Phi_0(\frac{\e_a}{K})^{1/2}$ is a rescaled voltage. 

\begin{figure}[ht]
    \centering
        \includegraphics[width=0.45\textwidth]{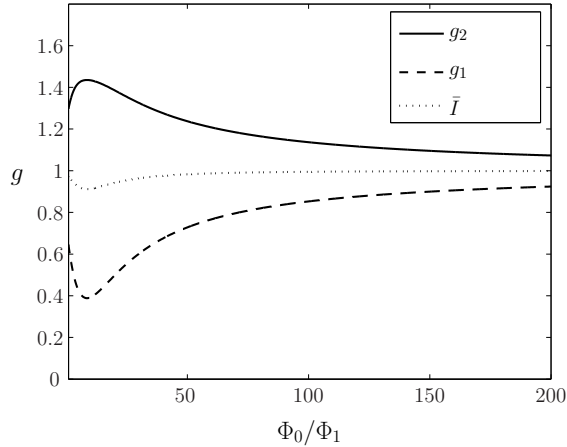}
    \caption{Dependence of gain of probe beam as a function of $\Phi_0/\Phi_1$. This is equivalent to increasing the external field. Note that the abscissa starts at 1, the lowest value for which our treatment is valid, although the difference between 1 and 0 is not visible on this scale.   Further structure may occur in the region $\Phi_0\Phi_1 \le 1$. 
     }
    \label{fig:g_nu}
\end{figure}

In the limit of high voltages, the $\kappa$ term in eq.(\ref{theta2}) can be neglected, and hence the reorientation $\theta(x,z)\propto \Phi_0^{-1}$. Hence at high voltages the director approaches a uniform distribution, in which case the beam-coupling disappears and $g=1$. In our approximation, the low voltage limit $\Phi \to 0$ is inaccessible; the minimum voltage for which our approximations are valid is $\Phi_0 \approx 1$. This follows because we have assumed that the modulated component of the electric field is small by comparison with the external field in the derivation of the liquid crystal director distribution eq.(\ref{eq1}).    The dependence of gain on the external potential is shown in Fig. \ref{fig:g_nu}. For $\Phi_0=1$  $g \approx 1.3$. The gain reaches a maximum $g \approx 1.44$ at $\Phi_0 \approx 8$. For $\Phi_0 >8$ the energy exchange parameter $g$ decreases monotonically toward a value of unity (i.e. no energy exchange) in the high potential limit.

%\clearpage

\section{Discussion and conclusions}

%\subsection{Conclusions}

In this paper we have carried out a model phenomenological calculation of energy exchange between beams incident on a thin liquid crystal grating sandwiched between two photoconducting layers. In this model calculation, the liquid crystal is subject to homeotropic boundary conditions, but this is not an essential feature of the model.  The energy exchange involves diffraction by an induced grating, with the exchange occurring when each incident wave is diffracted into the outgoing path of the other.  We find that there is a regime in which  significant energy exchange can occur, without leakage into other higher order diffracted waves. There is also another regime in which such leakage does occur.  

We find a maximal gain  of $g=1.45$, and this occurs for a grating wavelength of the order of the thickness of the sample. This appears to be a robust result, and qualitatively consistent with experiment \cite{khoo}. There is also significant dependence on the bulk voltage. In the limit of high voltage there is no effect (because there is no director modulation and hence no grating). As the voltage is reduced the effect increases. Our calculation does not permit the evaluation of the low-voltage limit, but we do find a maximum when the ratio of the external voltage to the surface modulation is about 10. Unfortunately the phenomenological nature of our calculation does not permit us to make any quantitative predictions with respect to actual voltage or beam intensities required to achieve this. However, the gain maximum as a function of voltage is also manifested as a maximum with respect to varying beam intensity. Although this calculation is vague with respect to quantitative prediction, we believe that the existence of a maximum as a function of voltage is a qualitatively robust result.

The calculation is broken down into a number of parts. Firstly we have supposed that interference between the incident beams affects the photoconducting layers by only changing the electric potential at the boundaries of the sample, and does so in proportion to some power of the beam intensities. Secondly we have calculated the modification of the  electric field inside the liquid crystal sample, supposing that there is a zeroth order field due to some imposed bulk potential. Thirdly, we have used the electric field to calculate the modulated director distribution, which necessarily then acts as an optical grating.  Fourthly, we have investigated the transmission of each beam through the modulated liquid crystal layer using a WKB-like approximation in the spirit of geometrical optics. The result of this calculation is a phase and amplitude optical profile for the extraordinary wave along the outgoing surface. Finally, using this surface optical profile we have used the Kirchoff method to evaluate the far field, and hence diffraction and inter-beam energy exchange. We may note that the resulting expression for the intensity of the  diffracted beams of different orders is reminiscent of the analogous calculation for diffraction through a thin grating in the Raman-Nath regime \cite{raman_nath}. However, the standard Raman-Nath calculation does not apply here.  The existence of inhomogeneities in the dielectric function perpendicular to the grating  direction complicates the calculation.  

Some features of our calculation are simplified in order to make the problem  tractable. We have measured the  potential induced at the surface with respect to the bulk voltage across the cell. An alternative low external voltage expansion would also in principle be possible, but we have not pursued this approach here. In addition, we have solved the liquid crystal director in a one-elastic-constant approximation,   and also linearized the Euler-Lagrange equations coupling the director and the electric potential. Neither of these approximations is essential, but there is significant potential advantages in obtaining   analytic formulas in what could otherwise be a computational minefield. The optical scattering in the system itself implies that the picture in which the incident beams penetrate the liquid crystal  layer unimpeded to provide potential modulations at the outgoing surface is only the first step in an iterative procedure. While it is possible to carry out this iteration in our model, we have chosen not to do so. This is partly because we would lose what analytic simplification we have achieved. Also,however, given that we only have a phenomenological model, we would in any case be no closer at this stage to a quantitative comparison with experiment.     

One particularly interesting feature of our calculations is that we are able to identify separately components of the refractive index modulation which are respectively in-phase  and  out-of phase with the mean optical field intensity. It is often stated that if the refractive index and optical field modulations are in-phase with respect to each other, then no two-beam coupling would be expected. However, notwithstanding the inaccuracies and approximations involved in our calculations, we find that this statement is unambiguously false. The in-phase beam coupling is indeed lower, but no by means identically zero.    

Although we have been able to obtain a semi-analytic form for the beam coupling, the   calculation is complicated. It involves electric fields, director distributions and light transmission through an inhomogeneous medium. The result is that the final magnitude of the effect under consideration seems to bear no simple relation to the rather large number of parameters which enter the problem. We can say definitively that the external voltage, the angle of incidence, the angle between the two beams, not to mention the thickness of the cell and the liquid crystal elastic constants, all play an important role, and furthermore the response is not monotonic. From an engineering point of view, there are clearly several possible ways to control the energy exchange process. 

But apart from the pronounced maximum in beam coupling when the grating width is of the order of the thickness of the sample, we are unable  at this stage to make further robust comments concerning the functional relationships without resorting to specific calculations. We cannot say whether further studies, and in particular a reliable microscopic theory, will clarify the situation. 

The theory presented in this paper can be developed in a number of ways. It can be trivially extended to liquid crystal cells in which only a single photoconductive layer attached. Alternatively, we might extend the present work, involving homeotropic surfaces, to low voltages, or to liquid crystal cells with homogeneous boundary conditions. Such a theory would be applicable, for example, to the experiments of Pagliusi and Ciparrone \cite{N15}. 
We also note that in this paper, the director distribution throughout the sample is clustered around the homeotropic direction.  However,  one might expect intuitively that the most dramatic effects would occur when the voltage modulation and the external field conspire to  produce large director shifts between one part of the sample and another. Our framework may permit such a calculation.   

The main weakness of the theory concerns the nature of the relationship between the potential modulations $\Phi_1$ and the beam intensities. The lack of relevant experimental data is partly because, as far as we are aware, the present paper is the first suggestion that the main mechanism for photorefractive beam coupling involves this quantity. We are hopeful that future work will therefore remedy this deficiency. An experiment which measures this quantity might involve Frederiks transition measurements in the presence of an externally applied optical beam.  

At a later stage, we would also hope to make contact between this theory and a more microscopic theory which elucidate processes in the photoconductive media, and at the photoconductive layer-liquid crystal interface. A second weakness involves the geometric optics approximation, and this restricts our calculations to the short wavelength limit. In most liquid crystal cells, this will be sufficient, but in principle longer-wavelength corrections are interesting. Indeed, we are currently carrying out optical calculations in which the optics is treated by solving the Maxwell equations exactly.
Such a scheme will automatically permit a self-consistent solution of the optics-potential-elastic problem.

\section*{Acknowledgements}
This work has been partially supported by a  Royal Society Joint
Project Grant ``Modelling the electro-optical properties of
ferroelectric nematic liquid crystal suspensions" awarded to TJS
and VYR (2003-5), a    NATO Grant CBP.NUKR.CLG.981968
``Electro-optics of heterogeneous liquid crystal systems"
coordinated by TJS (2006-8) and an  INTAS Young Scientist Fellowship
Award 1000019-6375 to VOK (2007-8). We also gratefully acknowledge
discussions with Dr. Malgosia Kaczmarek and Dr. Giampaolo
D'Alessandro (Southampton), Prof. Anatoli Khizhnyak (Metrolasers, California), and Prof. Ken Singer (Cleveland, USA). 
\bibliography{kubytskyi_et_al_nov07}

\end{document}